\documentclass[review]{elsarticle}

\usepackage{lineno,hyperref}
\usepackage{bbm}
\usepackage{pgfplots}
\usepackage{bigstrut}
\usepackage{booktabs,rotating,caption}
\usepackage{adjustbox}
\usepackage{longtable,lscape}
\newcommand{\hists}[1]{

\begin{tikzpicture}
    \begin{axis}[
        ymin=0, ymax=20,
        area style,xticklabels from table={#1}{x},
        xtick=data,
        xticklabel style = {rotate=90,anchor=east
        }
        ]
       \addplot+[ybar,mark=no,black,fill = green] table[x=z, y=y]{#1};
    \end{axis}
\end{tikzpicture}

}

\newcommand{\histsYear}[1]{

\begin{tikzpicture}
    \begin{axis}[
        ymin=0, ymax=5.5,
        area style,xticklabels from table={#1}{x},
        xtick=data,
        xticklabel style = {rotate=90,anchor=east
        }
        ]
       \addplot+[ybar,mark=no,black,fill = red] table[x=z, y=y]{#1};
    \end{axis}
\end{tikzpicture}

}

\journal{Computer Science Review}

\begin{document}

\begin{frontmatter}

\title{Problem-Space Evasion Attacks in the Android OS: a Survey}

\author{Harel Berger\corref{mycorrespondingauthor}}
\ead{harel.berger@msmail.ariel.ac.il}
\cortext[mycorrespondingauthor]{Corresponding author}
\author{Dr. Amit Dvir}
    \address{Ariel Cyber Innovation Center, Computer Science Department , Ariel University, 65 Ramat HaGolan street, Ariel, Israel}
\ead{amitdv@g.ariel.ac.il}

\author{Dr. Chen Hajaj}
\ead{chenha@ariel.ac.il}
\address{Ariel Cyber Innovation Center, Data Science and Artificial Intelligence Research
Center, Industrial Engineering and Management Department, Ariel University, 65 Ramat HaGolan street, Ariel, Israel}

\begin{abstract}
  Android is the most popular OS worldwide. Therefore, it is a target for various kinds of malware. As a countermeasure, the security community works day and night to develop appropriate Android malware detection systems, with ML-based or DL-based systems considered as some of the most common types. Against these detection systems, intelligent adversaries develop a wide set of evasion attacks, in which an attacker slightly modifies a malware sample to evade its target detection system. In this survey, we address problem-space evasion attacks in the Android OS, where attackers manipulate actual APKs, rather than their extracted feature vector. We aim to explore these kind of attacks, frequently overlooked by the research community due to a lack of knowledge of the Android domain, or due to focusing on general mathematical evasion attacks - i.e., feature-space evasion attacks. We discuss the different aspects of problem-space evasion attacks, using a new taxonomy, which focuses on key ingredients of each problem-space attack, such as the attacker model, the attacker's mode of operation, and the functionality assessment of post-attack applications.
\end{abstract}


\end{frontmatter}


\section{Introduction}
In March 2022, the Deep Instinct Threat Research published the 2022 Cyber Threat Landscape Report~\cite{2022threat}. This report describes an increase of 125\% in threat types and novel evasion techniques compared to 2021. The authors also stated that bad actors in the cyber world invest in the development of anti-AI and adversarial attacks and integrate these methods into their larger evasion strategy. In evasion attacks or techniques, one refers to a set of manipulations an attacker runs on a malicious sample, to evade detection by a target detection system. These manipulations are called \textit{evasion attacks}. Evasion attacks are devised in two ways: theoretical and physical. The theoretical way is referred to as a feature-space evasion attack, and the physical one is commonly referred to as a problem-space evasion attack. The targets of these attacks, detection machines that are ML-based or DL-based, do not analyze APKs or PEs, or any other type of malware. Instead, these detection systems require a mapping of the malware sample to numerical/textual feature vectors. In feature-space evasion attacks, an attacker manipulates the feature vector using various algorithms. Feature-space attacks are general because they utilize the representation of a sample and not the actual sample. On the other hand, problem-space evasion attacks use different manipulations on the actual sample. Problem-space evasion attacks are more complex to implement compared to feature-space evasion attacks, as they require a clear understanding of the subject domain~\cite{berger2020evasion,oakland2014,ndss2016,tong2019improving,hajaj2022less}. 

One of the popular environments for evasion attacks is the Android OS. As a consequence, an impressive amount of detection systems were devised in recent years for both malware and evasion attacks, as reported in~\cite{arshad2016android,odusami2018android,qiu2020survey,pan2020systematic,kouliaridis2021comprehensive,omer2021efficiency,kambar2022survey}. Most Android malware detection systems take one of three courses: static analysis, behavioral analysis, or dynamic analysis. Static analysis detection systems explore specific content from the files of the application, like permission requests or the sequences of API calls (e.g.,~\cite{arp2014drebin,onwuzurike2019mamadroid}). Other detection machines explore the execution of the application (e.g., ~\cite{wong2016intellidroid,yuan2014droid}). Behavioral analysis inspects CPU usage, the number of in-going and out-going network packets, etc. (e.g.,~\cite{shabtai2012andromaly,burguera2011crowdroid}). Hybrid machines fuse several approaches to detect malware (e.g.,~\cite{martinelli2017bridemaid,lu2020android}). Evasion attacks, and more specifically, problem-space evasion attacks, try to evade detection by these detection systems. Each attack targets different kinds of detection systems. 

The core contributions of this survey are threefold. First, we explore influential research and tools of problem-space evasion attacks against Android malware detection systems in recent years, between 2012 and 2022.  
Various surveys were conducted on problem-space evasion attacks, e.g.~\cite{ibitoye2019threat,li2021arms,ling2021adversarial,yuan2019adversarial,chakraborty2018adversarial,li2018security,liu2018survey,zhang2020adversarial,park2020survey,elsersy2022rise,de2020survey,pitropakis2019taxonomy,selvaganapathy2021review}. However, some of these surveys are general and therefore do not focus on a specific domain of evasion attacks~\cite{li2021arms,li2018security,liu2018survey,park2020survey,de2020survey,pitropakis2019taxonomy}. Other surveys analyze domains other than Android OS, such as adversarial network examples, Windows PEs, Natural Language Processing, and images~\cite{ibitoye2019threat,ling2021adversarial,yuan2019adversarial,chakraborty2018adversarial,zhang2020adversarial}. Several surveys focused on evasion attacks on the Android OS domain, for example~\cite{bhusal2022adversarial,elsersy2022rise,selvaganapathy2021review}. While Bhusal and Rastogi~\cite{bhusal2022adversarial} presented work on the nature of feature-space and problem-space attacks, they only explored one problem-space evasion attack. Also, ~\cite{elsersy2022rise,selvaganapathy2021review} surveys suggest a different taxonomy, which focuses on the evasion techniques, while our survey analyzes different aspects that have increased in recent years; For example, the importance of functional evaluation~\cite{berger2020evasion} and the orientation of different problem-space attacks.  Second, exploring problem-space attacks is vital to test the realistic assumptions of feature-space evasion attacks. For example, Berger et al.~\cite{berger2022you} showed that feature-space attacks in the Android domain do not serve as proxies for problem-space evasion attacks. In other words, feature-space attacks do not depict reality accurately. Therefore, problem-space attacks are great candidates to test the validity of feature-space attacks, as they are constructed through realistic changes to the application. Acknowledging different problem-space attacks is important as a way of testing existing and new evasion attacks for every domain, specifically for the Android OS domain. Finally, we present a new taxonomy of problem-space attacks using the attacker model, orientation, functionality assessment, and types of manipulation on the application. Note that every evasion attack presented in previous work is included in our survey as well. In total, this survey is presented for the community to aid future research in problem-space evasion attacks, and complete the assessment of evasion attacks on the Android OS domain. 

The remainder of this paper is organized as follows: First, the background on APKs and feature types of ML-based Android malware detection systems are described in Section~\ref{background}. Then, our taxonomy of problem-space evasion attacks is presented in Section~\ref{taxonomy}.
Next, a full discussion on insights from the explored works is given in Section~\ref{diss}. The paper is concluded in Section~\ref{conclusion} with suggestions for future research.

\section{Background}
\label{background}
This section presents the background of our survey. First, the APK structure in a nutshell in Section~\ref{apk_st}. Next, the types of ML-based Android malware detection systems in Section~\ref{feature_types}. These machines are targets of problem-space evasion attacks. Therefore, since this survey explores evasion attacks, it is important to clarify the categories of targets of these attacks. 
\subsection{APK File}
\label{apk_st}
The Android PacKage (APK) is the file format used by the Android application markets. APK is a compressed file containing the following files: \emph{manifest}, \emph{classes.dex}, \emph{layout files}, \emph{res}, and \emph{assets}.
The manifest file contains information that is essential for the APK, including the required user permissions. Another vital component of the APK is the binary code, classes.dex, which can be converted to several reverse engineering languages (e.g., Smali). Graphic resources are ordered on each page of the application using the layout files. Additional files which are not code files, like pictures or voice recordings, reside in res and assets directories. A more detailed explanation can be found in~\cite{struct_dev}. In this survey, we focus on the manifest file and the code files.
\subsection{Feature types of ML-Based Android Malware Detection}
\label{feature_types}
Several ML-based approaches were suggested to detect Android malware, which can be categorized into three main approaches. The first approach enumerates static information from the application, such as API calls or permission requests. This approach is termed static analysis~\cite{arp2014drebin,onwuzurike2019mamadroid,aafer2013droidapiminer,demontis2017yes,li2019android,li2020adversarial,berger2022mamadroid2,fereidooni2016anastasia,ou2022s3feature}. One of the most well-known Android malware detection systems using static analysis is Drebin~\cite{arp2014drebin,demontis2017yes,li2019android,li2020adversarial}, which gathers different types of information as features - permission requests, software/hardware components, intents, suspicious/restricted API calls, used permissions in the app's run, and URL addresses. Another famous detection machine that follows static analysis is MaMaDroid~\cite{onwuzurike2019mamadroid,berger2022mamadroid2}, which builds a control flow graph from the series of API calls from the code files (without any running of the application) and maps the transitions between the API calls. The second approach captures running of system calls while running the application. This approach is called dynamic analysis~\cite{bhatia2017malware,feng2018novel,alzaylaee2020dl,hou2016deep4maldroid,yuan2016droiddetector,alzaylaee2017emulator}. One of the famous works in this field is EnDroid~\cite{feng2018novel}, which analyzes system-level call traces and malicious
application-level behaviors. Deep4MalDroid~\cite{hou2016deep4maldroid} is another detection machine that uses dynamic analysis to create system call graphs. The third approach analyzes the behavior of the application using CPU or battery usage, network packets sent and received, and other parameters that describe the behavior of the application. This approach is termed behavioral analysis~\cite{shabtai2012andromaly,shabtai2010intrusion,shabtai2014mobile,saracino2016madam,wang2021android}. Andromaly~\cite{shabtai2012andromaly}, one of the well-known detection machines that analyze applications via behavioral analysis, explores network communication 
traffic patterns. A similar approach was presented in~\cite{shabtai2014mobile}. These detection systems enumerate the RTT values, the number of packages that were sent and received, etc. A hybrid approach combines multiple types of features from different systems~\cite{yuan2014droid,martin2019android,wang2015reevaluating,lindorfer2015marvin,ding2021hybrid}. A famous hybrid Android malware detection system was suggested by Martín et al.~\cite{martin2019android}. This system utilizes static and dynamic analysis of Android applications. In particular, the transitions between states of execution and of API calls are explored by this system. Marvin~\cite{lindorfer2015marvin}, another well-known hybrid Android malware detection system, focuses on permissions, certificates, etc. In addition, a dynamic analysis of several is processed, including phone activities like data leakages and network communication. This survey explores attacks against the three main approaches: static, dynamic, and behavioral.

\section{Taxonomy Characteristics and Table}
\label{taxonomy}
\begin{figure}[h!]
\includegraphics[width=0.85\textwidth]{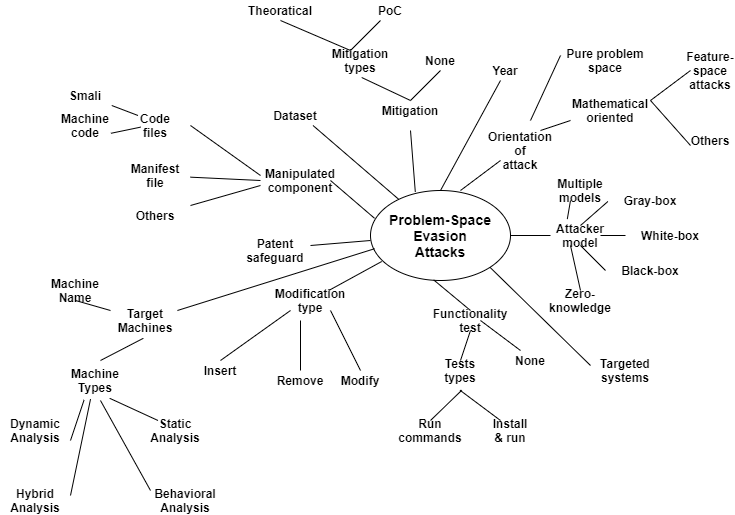}
\caption{Our Taxonomy of Problem-Space Evasion Attacks}
\label{fig:tax_pic}
\end{figure}

This section defines characteristics for the taxonomy of problem-space evasion attacks\footnote{The idea of this survey is to explore the quality of works on problem-space evasion attacks and the gaps that are still there in this type of research. Therefore, works that did not include an evaluation of target classifiers, such as Obfuscapk~\cite{aonzo2020obfuscapk} were eliminated from this survey. Such attacks and attack tools have not been proven to be efficient against any target. Also, works that discuss an attack but do not clearly describe an algorithm/process of the evasion attack were excluded as well (e.g.,~\cite{crussell2014andarwin,khanmohammadi2017hydroid}). Finally, works exploring existing manipulated benchmarks or obfuscation methods that were implemented in original applications from known datasets~\cite{cai2018droidcat,suarez2017droidsieve} were left out, as this survey follows evasion attacks on existing malware, not sophisticated malware datasets.}, that are presented in Table~\ref{attacks_table}. A graphic presentation of our taxonomy can be found in Fig.~\ref{fig:tax_pic}. Each of the following characteristics is described by its column name in the table. The explanation of each characteristic includes a set of appropriate values that were used in the table (if applicable).
\begin{itemize}
    \item \textbf{Targeted machines:} The name of the targeted machine/s (\textbf{Targets}) and the analysis type of this machine (\textbf{Tar. type}) - Static (\textbf{S}), Dynamic (\textbf{D}), or Behavioral (\textbf{H}).
    \item \textbf{Attacker model (At. Mod.):} the attacker model depicts the knowledge that the attacker has of the target detection machine. The options for the attacker models are White-box (\textbf{W}), Black-box (\textbf{Bl}), Gray-box (\textbf{G}) and Zero-knowledge (\textbf{Z}). Some studies used multiple attacker models (\textbf{Mu}) to test different types of attackers, with various capabilities. 
    \item \textbf{Orientation of the attack (Orient.):} Some of the work in this field was implemented through exploration of the application, generating perturbations, and examination of their effects on the target systems (\textbf{P}). Other works utilize general feature-space attacks or other mathematical abstractions and implement changes to the app according to the results of these abstractions (\textbf{F}). A similar split was suggested by Park et al.~\cite{park2020survey}, into two groups - gradient-driven or problem-driven. However, some of the works mentioned in our survey utilize other types of mathematical abstractions than the use of the gradient or even feature-space attacks. Therefore, we split the work into pure problem-space attacks or mathematically oriented attacks. 
    \item\textbf{Datasets:} The benign (\textbf{DS-Ben}) and malicious (\textbf{DS-Mal}) datasets that were used in each study. 
    \item \textbf{Manipulated component (Man. comp.)}: The concrete part of the app that was manipulated in the attack. Specifically, the optional parts are the manifest file (\textbf{M}), or the code files (\textbf{S}) - in their binary version or with their conversion to the Smali language - or both of these parts (\textbf{B}).
    \item \textbf{Modification type (Mod. type):} Changing a malicious app to evade classification can be done using one of the following: insertion of content (\textbf{I}), removal of content (\textbf{R}) or alteration of content (\textbf{A}).
    An example of the differences between the options takes the form of a permission request of SEND\_SMS. The attacker can add this request to an app (\textbf{I}), remove it (\textbf{R}), or change the occurrence of it in an app to READ\_SMS or simply SMS (\textbf{A}).
    \item \textbf{Detection decrease (Det. dec.):} The success of the attack against the targeted detection system. In other words, the decrease in detection rate between the original applications and the manipulated counterparts.
    \item \textbf{Mitigation techniques (Mit.):} Some of the studies suggested or implemented a proof-of-concept mitigation technique for the attack (\textbf{PoC}). Other studies described mitigation techniques in theory, without any implementation or evaluation (\textbf{T}). Other studies did not suggest any defense against the attacks described in the study (\textbf{N}). 
    \item \textbf{Patent safeguard (Pat.):} Some attacks utilize obfuscation techniques to evade classification of malicious applications as malicious. These techniques can be used by innovative developers of an application to hide their patents from unknown entities that want to analyze and copy them. This survey suggests which problem-space attacks can be referred to as a safeguard to patents as well, aside from finding weak spots in detection machines. The options of this characteristic are True (\textbf{V}), Partial (\textbf{Pa}) or False (\textbf{X}).
    \item \textbf{Functionality assessment (Func.):} As problem-space evasion attacks manipulate actual APKs, the modifications they carry out may harm the functionality of the application. It is important to test the functionality of a sample from the manipulated data to ensure that the attack does not create a non-operational app. The options of this characteristic are no functionality test (\textbf{N}), installation \& run process only (\textbf{IR}), and running commands (\textbf{C}).
\item \textbf{Year:} The year of publication, between 2012-2022.
\end{itemize}

\begin{landscape}
 \begin{longtable}{|p{2cm}|p{0.9cm}|p{0.8cm}|p{1.1cm}|p{0.6cm}|p{2.9cm}|p{0.8cm}|p{0.7cm}|p{0.7cm}|p{0.7cm}|p{0.7cm}|p{2.2cm}|p{2.2cm}|p{0.8cm}|}
 \caption{Our taxonomy of problem-space evasion attacks in Android OS}
\label{attacks_table}
\\\hline
\textbf{Name}      & \textbf{Man. comp.} & \textbf{Mod. type} & \textbf{Orient.} & \textbf{Fun.} & \textbf{Targets} & \textbf{At. mod.}   & \textbf{Det. Dec.}& \textbf{Mit.}& \textbf{Pat.} & \textbf{Tar. type} & \textbf{DS-Ben}   & \textbf{DS-Mal} & \textbf{Year} \\ \hline
\endhead

Pierazzi et al.~\cite{pierazzi2020intriguing}     & B                          & I                            & F            & IR                         & Drebin~\cite{arp2014drebin},\newline SecSVM~\cite{demontis2017yes}                                                                                                                                & W & 100                                   & T                     & X      & S              & Androzoo~\cite{allix2016androzoo}                                        & Androzoo~\cite{allix2016androzoo}                                                & 2020                      \\ \hline
Crystal Ball paper~\cite{berger2021crystal}       & M                            & A                            & P            & C                         & Drebin~\cite{arp2014drebin}, SecSVM~\cite{demontis2017yes}, FM~\cite{li2019android}, DNN~\cite{li2020adversarial}, VT~\cite{total2012virustotal}                                                                                                                   & Mu     & 100                                   & T                     & X      & S              & Androzoo~\cite{allix2016androzoo}                                        & Drebin~\cite{arp2014drebin}                                                  & 2021                      \\ \hline
Berger et al.~\cite{berger2020evasion}     & S                            & M                            & P            & C                         & Drebin~\cite{arp2014drebin}                                                                                                                                        & W          & 20                                    & N                     & V      & S              & Androzoo~\cite{allix2016androzoo}                                        & Drebin~\cite{arp2014drebin}                                                  & 2020                      \\ \hline
Android-HIV~\cite{chen2019android}        & S                            & I                            & F            & N                         & Drebin~\cite{arp2014drebin}, MaMaDroid~\cite{onwuzurike2019mamadroid}                                                                                                                             & Mu  & 97                                    & T                     & X      & S              & Drebin~\cite{arp2014drebin}, PlayDrone~\cite{viennot2014measurement}                              & Drebin~\cite{arp2014drebin}, VirusShare~\cite{virusshare}, Apkpure~\cite{apkpure}                             & 2021                      \\ \hline
SecSVM paper~\cite{demontis2017yes}       & S                            & A                            & F            & N                         & Drebin~\cite{arp2014drebin}, SecSVM~\cite{demontis2017yes}                                                                                                                                & Mu            & --              & PoC                     & Pa      & S              & Drebin~\cite{arp2014drebin}                                          & Drebin~\cite{arp2014drebin}, Contagio~\cite{Contagio}                                        & 2017                      \\ \hline
MaMaDroid-2.0~\cite{berger2022mamadroid2}         & S                            & A                            & P            & C                         & MaMaDroid~\cite{onwuzurike2019mamadroid}                                                                                                                                     & Mu     & 100                                   & PoC                     & X      & S              & Androzoo~\cite{allix2016androzoo}                                        & Drebin~\cite{arp2014drebin}                                                  & 2022                      \\ \hline
DroidCha-meleon~\cite{rastogi2013droidchameleon}          & B                        & A                            & P            & N                         & AVs from VT~\cite{total2012virustotal}                                                                                                                                   & Z             & 100                                   & T                     & Pa      & S              & Google-play market~\cite{gmarket}                               & Contagio~\cite{Contagio}                                                & 2013                      \\ \hline
ADAM~\cite{zheng2012adam}               & S                            & A                            & P            & N                         & AVs from VT~\cite{total2012virustotal}                                                                                                                                   & Z             & 73+ & T                     & Pa      & S              & -                                               & Antiy~\cite{antiy},\newline Contagio~\cite{Contagio}                                              & 2012                      \\ \hline
MRV~\cite{yang2017malware}                & S                            & I                            & F            & N                         & AppContext~\cite{yang2015appcontext}, Drebin~\cite{arp2014drebin}                                                                                                                            & Bl         & 60                                    & PoC                     & X      & S              & Google-play market~\cite{gmarket}                               & Genome~\cite{zhou2012dissecting}, Contagio~\cite{Contagio}, VirusShare~\cite{virusshare}, Drebin~\cite{arp2014drebin}, Google-play market~\cite{gmarket} & 2017                      \\ \hline
Pomilla et al.~\cite{pomilia2016study}      & B                          & A                            & P            & N                         & AVs from VT~\cite{total2012virustotal}                                                                                                                                   & Z             & 60+ & N                     & Pa      & S              & -                                               & Contagio~\cite{Contagio},\newline Drebin~\cite{arp2014drebin},\newline Andro-total                              & 2016                      \\ \hline
Canfora et al.~\cite{canfora2015obfuscation} & B         & M,I       & P         & N         & VT~\cite{total2012virustotal}        & Z      & 100 & N         & Pa         & S         & -                 & Drebin~\cite{arp2014drebin}                       & 2015 \\ \hline
Aydogan et al.~\cite{aydogan2015automatic}      & S                            & A                            & P            & N                         & AVs from VT~\cite{total2012virustotal}                                                                                                                                   & Z             & --               & N                     & V      & S              & -                                               & Genome~\cite{zhou2012dissecting}                                               & 2015                      \\ \hline
Wang et al.~\cite{DBLP:journals/corr/abs-2111-10085}         & B                          & I                            & F            & N                         & AVs from VT~\cite{total2012virustotal},   and Drebin~\cite{arp2014drebin} (LGBM, SVM and RF, DNN)                                                                                             & Mu & 100                                   & Pa                     & X      & S              & Drebin~\cite{arp2014drebin}                                          & Drebin~\cite{arp2014drebin}                                                  & 2022                      \\ \hline
Cara et al.~\cite{cara2020feasibility}         & S                            & I                            & F            & N                         & (MLP)                                                                                                                                         & Mu      & 100        & N                     & X      & S              & Androzoo~\cite{allix2016androzoo}                                        & VT~\cite{total2012virustotal}, Drebin~\cite{arp2014drebin}, Contagio~\cite{Contagio}                            & 2020    \\ \hline
HRAT~\cite{zhao2021structural}               & S                            & A                            & F            & C                         & Malscan~\cite{wu2019malscan}, MaMaDroid~\cite{onwuzurike2019mamadroid}, APIgraph~\cite{zhang2020enhancing}                                                                                                                  & W & 100                                   & PoC                     & X      & S              & Malscan~\cite{wu2019malscan}                                         & Malscan~\cite{wu2019malscan}                                                 & 2021                      \\ \hline
Abaid et al.~\cite{abaid2017quantifying}        & S                            & I,R                          & P            & N                         & Drebin~\cite{arp2014drebin}                                                                                                                                        & Mu            & 100                                   & T                     & Pa      & S              & Drebin~\cite{arp2014drebin}                                          & Drebin~\cite{arp2014drebin}                                                  & 2017                      \\ \hline
Li et al.~\cite{li2021framework}           & S                            & I, R                         & F            & IR                         & Drebin-DNN~\cite{li2020adversarial}                                                                                                                                    & Mu            & 100                                   & PoC                     & X      & S              & Drebin~\cite{arp2014drebin}                                          & Drebin~\cite{arp2014drebin}                                                  & 2021                      \\ \hline
Mairoca et al.~\cite{maiorca2015stealth}      & B                          & A                            & P            & N                         & AVs from VT~\cite{total2012virustotal}                                                                                                                                   & Z             & 50+ & T                     & Pa      & S              & -                                               & Genome~\cite{zhou2012dissecting}, Contagio~\cite{Contagio}                                     & 2015                      \\ \hline
Pandora~\cite{protsenko2013pandora}            & B                          & A                            & P            & N                         & AVs from VT~\cite{total2012virustotal}                                                                                                                                           & Z             & 85                                    & N                     & Pa      & S              & -                                               & Mobile Sandbox dataset~\cite{spreitzenbarth2013mobile}                        & 2013                      \\ \hline
Evade-Droid~\cite{bostani2021evadedroid}         & S                            & I                            & F            & IR                         & Drebin~\cite{arp2014drebin}, SecSVM~\cite{demontis2017yes}, MaMaDroid~\cite{onwuzurike2019mamadroid}, ADE-MA~\cite{li2020adversarial}                                                                                                            & Bl         & 81                                    & T                     & X      & S              & Androzoo~\cite{allix2016androzoo}                                        & Androzoo~\cite{allix2016androzoo}                                                & 2022                      \\ \hline
Chen et al.~\cite{chen2019can}         & B                          & I                            & F            & N                         & DroidAPIminer~\cite{aafer2013droidapiminer}, Drebin~\cite{arp2014drebin}, Stormdroid~\cite{chen2016stormdroid}, and MaMaDroid~\cite{onwuzurike2019mamadroid}                                                                                             & W         & 80                                    & N                     & X      & S              & Google-play market~\cite{gmarket}                               & Genome~\cite{zhou2012dissecting}, Drebin~\cite{arp2014drebin},\newline Contagio~\cite{Contagio},\newline Pwnzen~\cite{pwnzen}                           & 2019                      \\ \hline
Vidas et al.~\cite{vidas2014evading}         & S                            & A                            & P            & C                         & Andrubis~\cite{lindorfer2014andrubis}, SandDroid~\cite{debelo2013sandroid}, Foresafe~\cite{forsafe}, Copperdroid~\cite{tam2015copperdroid}, AMAT~\cite{amat}, Mobile Sandbox~\cite{spreitzenbarth2013mobile}, Bouncer~\cite{bouncer}                                                                     & Z             & --               & T                     & X      & D/H              & -                                               & -                                                       & 2014                      \\ \hline
Dadidroid paper~\cite{ikram2019dadidroid}    & S                            & A                            & P            & N                         & MaMaDroid~\cite{onwuzurike2019mamadroid}, Dadidroid~\cite{ikram2019dadidroid}                                                                                                                          & Z             & 77                                    & PoC                     & Pa      & S              & Marvin~\cite{lindorfer2015marvin},\newline OldBenign~\cite{onwuzurike2019mamadroid},\newline NewBenign~\cite{onwuzurike2019mamadroid},\newline ObDataII~\cite{lindorfer2015marvin},\newline Packed-Apps~\cite{dong2018understanding} & Marvin~\cite{lindorfer2015marvin},\newline Drebin~\cite{arp2014drebin},\newline ObDataI~\cite{garcia2015obfuscation},\newline ObDataII~\cite{lindorfer2015marvin},\newline Packed-Apps~\cite{dong2018understanding}             & 2019                      \\ \hline
Hammad et al.~\cite{hammad2018large}       & B                          & A                            & P            & C                         & AVs from VT~\cite{total2012virustotal}                                                                                                                                   & Z             & 100                                   & N                     & P      & S              & Androzoo~\cite{allix2016androzoo}/\newline Google-play market~\cite{gmarket}                             & Genome~\cite{zhou2012dissecting}, Contagio~\cite{Contagio}, AndroTotal~\cite{maggi2013andrototal}, Drebin~\cite{arp2014drebin}, VirusShare~\cite{virusshare}        & 2018                      \\ \hline
Faruki et al.~\cite{faruki2014evaluation}  & B         & M,I       & P         & N         & VT~\cite{total2012virustotal}        & Z      & 100 & N         & Pa         & S         & Google-play market~\cite{gmarket} & Contagio~\cite{Contagio}, Genome~\cite{zhou2012dissecting}, VirusShare~\cite{virusshare} & 2014  \\\hline
Mystique~\cite{meng2016mystique}           & B                        & A                            & F            & N                         & Drebin~\cite{arp2014drebin}, Adagio~\cite{gascon2013structural}, Allix et al.~\cite{allix2014machine}, RevealDroid~\cite{garcia2015obfuscation}, ScanDroid~\cite{fuchs2009checking}, FlowDroid~\cite{arzt2014flowdroid}, DroidSafe~\cite{gordon2015information}, ICCTA~\cite{li2015iccta}, TaintDroid~\cite{enck2014taintdroid}, VT~\cite{total2012virustotal}                & Z             & 100                                   & T                     & Pa      & S              & Google-play market~\cite{gmarket}                               & Genome~\cite{zhou2012dissecting}                                                  & 2016                      \\ \hline
Mystique-S~\cite{xue2017auditing}        & B                          & A                            & F            & C                         & Droidbox~\cite{lantz2011droidbox}, Drozer~\cite{drozer},\newline Taintdroid~\cite{enck2014taintdroid}                                                                                                                 & Z             & 80                                    & N                     & X      & D/H              & Google-play market~\cite{gmarket}                               & Genome~\cite{zhou2012dissecting}                                                  & 2017                      \\ \hline
Divide-\&-Conquer~\cite{maier2014divide} & S                            & A                            & P            & C                         & Andrubis~\cite{lindorfer2014andrubis},\newline BitDefender~\cite{pavel2013bitdefender}, \newline ForeSafe~\cite{forsafe},Joe Sandbox Mobile~\cite{joes}, Mobile Sandbox~\cite{spreitzenbarth2013mobile},Sand-Droid~\cite{debelo2013sandroid},\newline TraceDroid~\cite{TraceDroid},\newline Trend Micro~\cite{Trend_micro}                                             & Z             & --               & T                     & X      & D/H              & -                                               & -                                                       & 2014                      \\ \hline
Grosse et al.~\cite{grosse2017adversarial}       & M                            & I                            & F            & N                         & Drebin-DNN~\cite{li2020adversarial}                                                                                                                                    & Bl         & 69                                    & PoC                     & X      & S              & Drebin~\cite{arp2014drebin}                                          & Drebin~\cite{arp2014drebin}                                                  & 2017                      \\ \hline
IagoDroid paper~\cite{calleja2018picking}          & S                            & I                            & F            & N                         & RevealDroid~\cite{garcia2015obfuscation}                                                                                                                                   & Mu & 97                                    & PoC                     & X      & S              & Drebin~\cite{arp2014drebin}                                          & Drebin~\cite{arp2014drebin}                                                  & 2018                      \\ \hline
Petsas et al.~\cite{petsas2014rage}       & S                            & A                            & P            & C                         & DroidBox~\cite{lantz2011droidbox}, DroidScope~\cite{yan2012droidscope}, TaintDroid~\cite{enck2014taintdroid}, Andrubis~\cite{lindorfer2014andrubis}, SandDroid~\cite{debelo2013sandroid}, ApkScan~\cite{apkscan}, VisualThreat~\cite{visual_threat}, Tracedroid~\cite{TraceDroid}, CopperDroid~\cite{tam2015copperdroid}, Apk Analyzer~\cite{apk-analyzer}, ForeSafe~\cite{forsafe}, Mobile Sandbox~\cite{spreitzenbarth2013mobile} & Z             & --               & T                     & X      & D/H              & -                                               & Contagio~\cite{Contagio}                                                & 2014                      \\ \hline
UAP~\cite{DBLP:journals/corr/abs-2102-06747}                & S                            & I                            & F            & N                         & Drebin~\cite{arp2014drebin}                                                                                                                                        & W         & 100                                   & N                     & X      & S              & Drebin~\cite{arp2014drebin}                                          & Drebin~\cite{arp2014drebin}                                                  & 2022                      \\ \hline
\end{longtable}

\end{landscape}

\section{Discussion}
\label{diss}
Table~\ref{attacks_table} describes our taxonomy on problem-space evasion attacks in the Android OS domain. This section describes the insights gained by viewing this table. Some of these insights describe correlations between different aspects, like modification types and attack orientation (Section~\ref{mod_attack_cor}), the attacker models and the types of attacks (Section~\ref{atmod_tyat}), or the patent safeguard and the orientation of the attack (Section~\ref{pat_ori}). Other insights are the result of the distribution of other aspects, such as the types of targeted machines (Section~\ref{tar_types}), the functionality assessments in the surveyed papers (Section~\ref{func_inst}), the manipulated component (Section~\ref{manip_comp}), years of publication (Section~\ref{years_dis}), and the datasets (Section~\ref{datasets_dis}).
\subsection{Modification Types and Attack Orientation}
\label{mod_attack_cor}
Most of the attacks, and specifically the mathematical-oriented ones, only insert lines into the code. This is a precaution for these types of attacks, as they do not integrate knowledge on the effects of the types of change on the functionality of the application. However, some of the cases show that these additions are too predictable and, therefore, may be mitigated easily. For example, Android-HIV~\cite{chen2019android} adds no-ops against the MaMaDroid detection machine~\cite{onwuzurike2019mamadroid}. Enumeration of no-ops can be done automatically. Consequently, this attack can be mitigated efficiently, as suggested by~\cite{berger2022mamadroid2}. An exception is found in the rewiring attack from HRAT ~\cite{zhao2021structural}, which modifies the flow of caller and callee functions inside an application to evade the same detection machine, MaMaDroid. On the other hand, most of the pure problem-space attacks modify the application. This mostly exemplifies the main idea that we mentioned earlier, problem-space evasion attacks that are based on mathematical abstractions are too general and therefore create a smaller threat to the security community than pure problem-space attacks.  
\subsection{Attacker Models and Types of Attacks}
\label{atmod_tyat}
Attacker models seem to be correlative to the types of attacks in general – attacks that follow the zero-knowledge model mostly do not correlate to mathematical-oriented problem-space attacks. Attacks that manipulate the app systematically based on its structure and content, even if they target a specific set of features to evade, do not require additional knowledge (e.g.,~\cite{berger2022mamadroid2,maier2014divide}). A more general approach, such as mathematical analysis of the feature set, requires more knowledge in practice to run the attack, such as the attack of Pierazzi et al.~\cite{pierazzi2020intriguing} which elevated the perfect knowledge/ White-box attacker model.
\subsection{Patent Safeguard and Attack Orientation}
\label{pat_ori}
Some of the attacks that we surveyed can be fully / partly used, as well as safeguard information or patents of the creator of the apps. Other attacks change parts of the application to evade detection, but cannot be considered as methods of patent safeguard. It is interesting to see that there is a correlation between this parameter and the orientation of the attack. Eighty-six percent of the attacks that were marked to be potentials for patent safeguard are pure problem-space evasion attacks, and only 14\% of the patent safeguard potentials are mathematical-oriented problem-space evasion attacks. On the other side, 74\% of the attacks that have no potential to aid app developers are of mathematical-oriented origin, and only 16\% of these attacks are pure problem-space attacks.

\subsection{Target Machine Types}
\label{tar_types}
Most of the problem-space evasion attacks that we surveyed are against static analysis detection machines. These targets are easier to explore by viewing their specific content from the app and trying to conceal it, as presented by the attacks against SecSVM and Drebin~\cite{demontis2017yes} or the Pandora framework~\cite{protsenko2013pandora}. Obfuscation of several API calls to seem like other API calls requires some code lines. On the other hand, changing an application to evade dynamic analysis detection systems, that follows the system calls, or behavioral feature of behavioral analysis detection systems requires full expertise of the Android OS and its functionalities, like the works of Petsas et al.~\cite{petsas2014rage} or Vidas et al.~\cite{vidas2014evading}. Therefore, most of the works cover attacks only against static analysis types of detection systems.
\subsection{Functionality Assessment}
\label{func_inst}
Most attacks do not include functionality evaluation. As explored in~\cite{berger2020evasion} – functionality assessment is critical to demonstrate, as the common belief that the manipulations that are carried out do not damage the functionality of the app is not enough in practice. Most of the mathematical-oriented and some of the pure problem-space attacks do not test their manipulated apps due to this mentioned common belief. The extent of functionality assessment is not defined and may not be defined at all – as some will say that \textit{X} apps are enough, and others will say that at least \textit{Y>X} apps should be investigated. However, some apps should be investigated. Moreover, a functionality test should include not only installing and running the app but carrying out some automatic clicks and touches on the screen, e.g. MonkeyRunner~\cite{monkey}, DroidBot~\cite{li2017droidbot}, GroddDroid~\cite{abraham2015grodddroid}, CuckooDroid~\cite{cuckoo}  or other tools. Some of these tools cover more ground, as they can be used as dynamic analysis tools for Android apps. However, as they run automatic operations on Android samples, they create a suitable environment for the functionality assessment as well. A standardization of functionality assessment is advised, to create an agreed platform in the community. This environment should be maintained constantly, to create a long-lasting tool to test application functionality\footnote{One of these tools, CuckooDroid~\cite{cuckoo}, was tested by us and found to be non-functional. We contacted the author of the tool and several colleagues regarding the operation of CuckooDroid, but it seems it is outdated.}.
\subsection{Manipulated Component}
\label{manip_comp}
Most attacks target code files or both code files and the manifest file. There are not many attacks that focus on the manifest file only, as it is a small environment for attacks. However, some works that solely target the manifest file show that manipulation of this file is no less effective~\cite{berger2021crystal,grosse2017adversarial}. 

\subsection{Publication Years}
\label{years_dis}
\begin{figure}
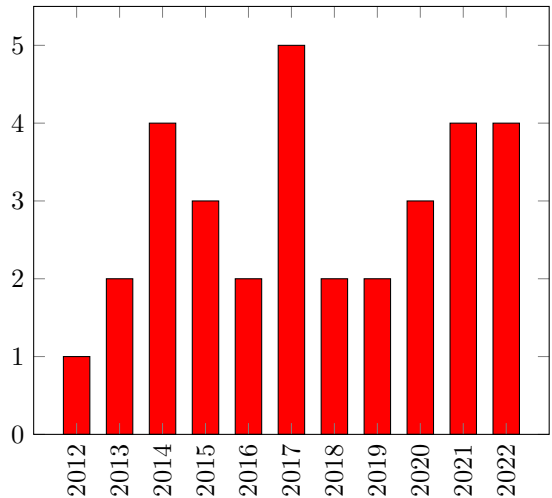

\centering
\histsYear{years.txt}
\caption[Distribution of problem-space evasion attacks on Android OS domain between 2012-2022]{Distribution of problem-space evasion attacks on Android OS domain between 2012-2022. \textbf{No color is needed.}}
\label{years_dis_fig}
\end{figure}

In this survey, we explored publications on the topics of problem-space evasion attacks in the Android OS domain between the years 2012 and 2022. As can be seen in Table~\ref{attacks_table} and also in Fig.~\ref{years_dis_fig}, every couple of years there is a small surge in publications in this area. The years 2014, 2017, and 2021 indicate more popularity of problem-space attacks in the Android OS domain.

\subsection{Datasets}
\label{datasets_dis}
\begin{figure}
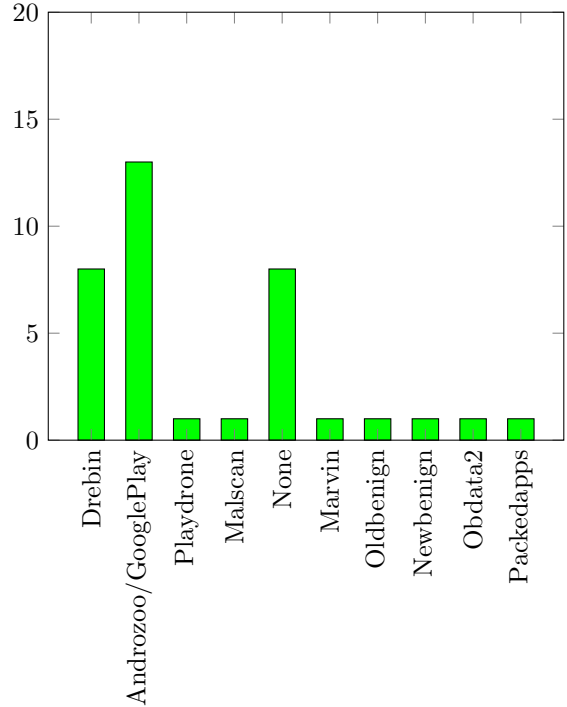

\centering
\hists{benapps.txt}
\caption[Benign datasets utilization in works on problem-space evasion attacks  ]{Benign datasets utilization in works on problem-space evasion attacks. \textbf{No color is needed.} }
\label{ben_apps_data_hist}
\end{figure}

\begin{figure}
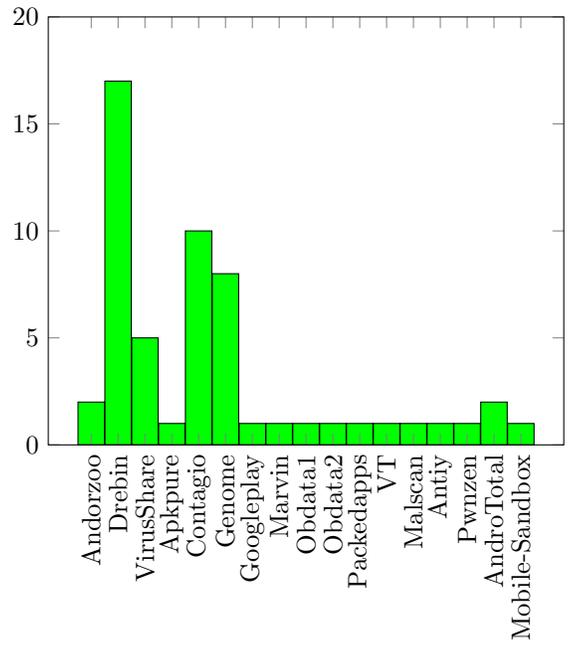

\centering
\hists{malapps.txt}
\caption[Malicious datasets utilization in works on problem-space evasion attacks ]{Malicious datasets utilization in works on problem-space evasion attacks. \textbf{No color is needed.}}
\label{mal_apps_data_hist}
\end{figure}

Different datasets were used to generate evasion attacks. To evaluate the success rates of these attacks, benign data is used as well. Figs.~\ref{ben_apps_data_hist}-\ref{mal_apps_data_hist} depict the distributions of datasets among the surveyed works. The most popular benign dataset in these works is Androzoo~\cite{allix2016androzoo}/GooglePlay market~\cite{gmarket}\footnote{These two sources are considered together, as some studies take apps from both, and as Androdzoo includes an ability to download apps from GooglePlay.}. Androzoo is a framework for downloading many applications from different years, and of different categories -  both benign and malicious applications. As Androzoo is updated constantly, no guarantee is given that each work in our survey that uses it as its source for benign apps chooses the same applications as the other works that do so unless each work publishes its hash values of each benign app (which does not happen in reality). The most popular malicious dataset in these works is Drebin~\cite{arp2014drebin}. Although the Drebin dataset is outdated, since it was gathered between 2010 and 2012, it has still been used in recent years. So far, no agreed dataset, benign or malicious, has been announced in the community. Throughout the years, several attempts to create a benchmark dataset for evaluations of Android malware detection were originated (i.e.~\cite{lashkari2018toward,liu2020androzooopen,wang2019rmvdroid}). However, they did not succeed, as proved in this survey and also in other surveys (e.g.,\cite{elsersy2022rise}). Consequently, a benchmark dataset of Android benign \& malicious apps is still needed.

\section{Conclusion and New Directions}
\label{conclusion}
This survey explored problem-space evasion attacks on Android OS between the years 2012 and 2022 from a bird's-eye view. The main idea of our survey was to find out the missing or incomplete information in past research in this field. Therefore, our survey explored parameters that other surveys put aside, such as the orientation of the attack (pure or mathematical oriented), the functionality assessment of the attack, the modification type, and the manipulated component inside the app. These parameters shed light on some interesting insights that skulk in the shadows, such as the correlations between attacker models and types of attack or the connections between modification types and attack orientation. The explored insights can be considered as pointers for the following future directions of research:
\begin{enumerate}
    \item \textit{Generation of new problem-space evasion 
attacks whose orientation is mathematical abstraction:} This survey found out that most of the mathematical oriented problem-space attacks utilize addition or removal of content, not actual alteration of the data. New research should investigate this aspect and try to find new attacks that follow this path. Of course, these attacks should be tested as to whether they create functional applications. Also, the strong connection of these attacks to the White-box, Gray-box, and Black-box attacker models leads to an interesting debate on whether new mathematical oriented problem-space evasion attacks can be generated using a zero-knowledge attacker model. 
\item \textit{Generation of new problem-space evasion attacks that target behavioral, dynamic, and hybrid detection machines:} This direction aims to find new problem-space attacks to tackle detection machines that are not from the static analysis type. Most of the problem-space attacks that were surveyed targeted detection systems from the static analysis type. Researchers are urged to devise new attacks against the other types of detection systems - dynamic analysis, behavioral analysis, and hybrid. 
\item \textit{Functionality assessment of problem-space attacks:} This survey found that most of the problem-space evasion attacks do not validate the functionality of the manipulated applications. Future research both on existing problem-space attacks and new ones should include functionality assessment. Also, standardization of this assessment should be discussed and agreed upon in the community.
\item \textit{Manipulated component inside the app:} This survey found that most of the problem-space evasion attacks target the code of the application (binary or Smali code files), and not other files like the manifest file. New studies should investigate the impact of significant modification of the manifest file, as was suggested by~\cite{berger2021crystal,grosse2017adversarial}.  
\item \textit{Dataset sparsity:} As too many studies used different datasets for their evaluation, no benchmark dataset was found in this survey. The community is advised to gather an agreed on dataset for the evaluation of new problem-space attacks and also for the new generation of Android malware detection systems.
\end{enumerate}

\section*{Acknowledgments}
This work was supported by the Ariel Cyber Innovation Center in conjunction with the Israel National Cyber Directorate of the Prime Minister's Office.

\bibliography{elsarticle-template}

\end{document}